\documentclass[onecolumn,preprintnumbers,amsmath,amssymb]{revtex4}

\usepackage{graphicx}
\usepackage{graphics}
\usepackage{dcolumn}
\usepackage{bm}

\def\BEq{\begin{equation}}
\def\EEq{\end{equation}}
\def\BEqA{\begin{eqnarray}}
\def\EEqA{\end{eqnarray}}
\def\BEn{\begin{enumerate}}
\def\EEn{\end{enumerate}}
\def\BWT{\begin{widetext}}
\def\EWT{\end{widetext}}
\def\a{\alpha}

\def\b{\beta}

\def\g{\gamma}

\def\s{\sigma}

\def\t{\tau}

\begin{document}


\title{Single-step controlled-NOT logic from any exchange interaction}

\author{Andrei Galiautdinov}
 \email{ag@physast.uga.edu}
\affiliation{
Department of Physics and Astronomy,
University of Georgia, Athens, Georgia
30602, USA
}

\date{\today}

\begin{abstract}
A self-contained approach to studying the unitary evolution of coupled 
qubits is introduced, capable of addressing a variety of physical 
systems described by exchange
Hamiltonians containing Rabi terms. The method automatically 
determines both the Weyl chamber steering trajectory and the accompanying 
local rotations. Particular attention is paid to the case of anisotropic 
exchange
with tracking controls, which is solved analytically. It is shown that, 
if computational subspace is well isolated,
any exchange interaction can always generate high-fidelity, single-step 
controlled-NOT (CNOT) logic, provided that both qubits can be 
individually manipulated. The results are then
applied to superconducting qubit architectures, for which several CNOT 
gate
implementations are identified. The paper concludes with consideration 
of two CNOT gate designs having high efficiency and operating with no 
significant leakage to 
higher-lying non-computational states.
\end{abstract}


\maketitle

\section{\label{sec:INTRO}Introduction}

Controllability of quantum mechanical systems has been the subject 
of numerous investigations in the last several years 
\cite{WEAVER2000, SOLOMON2001a,
KHANEJAPRA63, SOLOMON2001b, AKULIN2001, RABITZ2001a, KHANEJACHPH267, 
RAMAKRISHNA2001, RABITZ2001b, CLARK2005, SUGNY2007}. An important 
contribution by Khaneja et al. on time-optimal control 
\cite{KHANEJAPRA63, KHANEJACHPH267} has led to the development 
of rf-pulse sequences for NMR-spectroscopy with nearly ideal 
performance \cite{KHANEJA_JMR179}. In Refs. 
\cite{KHANEJAPRA63} and \cite{KHANEJACHPH267} it was assumed that the local 
terms in the Hamiltonian can be made arbitrarily large, which 
would allow an almost instantaneous execution of
single-qubit operations. However, such hard control mechanism is 
not applicable to quantum computing architectures based on 
superconducting Josephson devices, in which the relevant 
computational subspace must be kept well isolated at all times.

In this regard, the work of Zhang et al. on geometric 
theory of nonlocal two-qubit operations \cite{ZHANG, ZHANG1} 
acquires special significance. 
The authors introduced a convenient, 
geometrically 
transparent description of SU(4) local equivalence classes and 
then used it to develop several implementations of quantum logic 
gates that did not involve hard-pulse control sequences. The 
description of entangling operations presented in 
Refs. \cite{ZHANG} and \cite{ZHANG1} is based on the fact 
\cite{KHANEJAPRA63, KHANEJACHPH267} 
that any two-qubit quantum gate
$U\in {\rm U(4)}$ can always be written as a product, 
called the {\it Cartan 
decomposition},
\begin{eqnarray}
	\label{eq:cartandec}
	U = e^{i\varphi}k_1U_{\rm ent}k_2, \quad k_1, k_2 
	\in {\rm SU(2)}\otimes 
	{\rm SU(2)},
\end{eqnarray}
with
\begin{eqnarray}
	\label{eq:ent}
	U_{\rm ent} = e^{-(i/2)(c_1\sigma_1^x\sigma_2^x+
	c_2\s_1^y\s_2^y+
	c_3\s_1^z\s_2^z)}.
\end{eqnarray}
The triplet of numbers $\vec{c} = (c_1, c_2, c_3)$ in 
Eq. (\ref{eq:ent}) may be taken to represent the local 
class of $U$. 
In general, such representation is 
{\it not} unique due to the presence of symmetries 
mapping class 
vectors to other class vectors of the same equivalence 
class. However, 
it was
shown in \cite{ZHANG} that the correspondence {\it can} 
be made unique 
if $\vec{c}$ is restricted to a tetrahedral region of 
$\Re^3$, called 
a Weyl chamber. One such chamber is chosen to be 
canonical. It is described by the following three conditions 
\cite{ZHANG, TUCCI}:
\begin{description}
\item [(i)] $\pi > c_1 \geq c_2 \geq c_3 \geq 0$,
\item [(ii)] $c_1+c_2 \leq \pi$,
\item [(iii)] if $c_3 =0$, then $c_1\leq \pi/2$.
\end{description}

When a physical system evolves under the action of its 
Hamiltonian, $\vec{c}$ 
traces a trajectory inside the Weyl chamber, which explicitly 
shows the 
(continuous) sequence of dynamically generated local equivalence 
classes.
For example, the Hamiltonian $H=g\s_1^x\s_2^x/2$ generates the 
straight line 
$\vec{c}(t)=(gt,0,0)$ in the Weyl chamber. In {\it this} case, 
Eq. (\ref{eq:cartandec}) 
reduces to
\BEq
U(t)= e^{-itH}\equiv U_{\rm ent}(t), \quad k_1, k_2=1.
\EEq
After steering for a time $t_{\rm CNOT} = \pi/2g$ the system hits 
the gate
\BEq
\label{eq:CNOTMATRIX}
U(t_{\rm CNOT}) =
\frac{1}{\sqrt{2}}
\begin{pmatrix} 1 & 0 & 0 & -i \cr 0 & 1 & -i & 0 
\cr 0 & -i & 1& 0 \cr -i & 0 & 0 & 1 \end{pmatrix}, 
\EEq
with
\BEq
\vec{c}=\pi/2 \times (1,0,0),
\EEq
belonging to the controlled-NOT equivalence class. 
By flanking $U(t_{\rm CNOT})$ with additional local rotations 
$K_1$ and $K_2$, 
any 
gate in that class can be made. For example, to make the 
{\it canonical} CNOT gate we can take
\BEqA
		\label{eq:trueCNOTMATRIX}
		{\rm CNOT} =
		e^{i\,\pi/4}
		\underbrace{e^{-i(\pi/4)\, \s_1^y}\,
		   e^{i(\pi/4)\left(\s_1^x - \s_2^x\right)}}_{K_1}
		   U(t_{\rm CNOT})  \underbrace{e^{i(\pi/4)\s_1^y}}_{K_2}.
\EEqA

When Rabi terms are present in the Hamiltonian, the steering 
trajectory is no longer a straight line. In Ref. \cite{ZHANG}, 
the trajectory $\vec{c}(t)$ was calculated using the relation
between the class vectors and the local invariants \cite{MAKHLIN}. 
That method was applied in Ref. \cite{PLOURDE} to a CNOT gate design 
for flux-qubits with SQUID-based controllable coupling.

In the present paper we propose an alternative approach to finding 
the steering trajectory that does not rely on local invariants. Our 
goal is to develop a systematic procedure for calculating the 
entangling part $U_{\rm ent}(t)$ of the time-dependent gate $U(t)$ 
together with the accompanying it local rotations $k_1(t)$ and 
$k_2(t)$, so that
the Cartan decomposition (\ref{eq:cartandec}) could be determined
at every 
step of system's evolution. It turns out that due to a special property 
of the relevant to our problem generators of su(4) --- the closure 
under commutation and the existence of a central element --- the local 
rotations required to implement (\ref{eq:cartandec}) can be chosen in 
a particularly simple form, which mimics the form of the local Rabi 
parts of system's Hamiltonian. Due to such simplifying form of $k_1$ 
and $k_2$, the full problem of steering can be analytically solved in 
the experimentally important case of tracking control.

In the mathematical portions of this paper we will use the notation 
that is convenient for Lie algebraic manipulations:
\BEqA
	X_k=(i/2) \s_k^x, \;
	 \; XX=(i/2)\s_1^x\s_2^x, 
	 \; YY=(i/2)\s_1^y\s_2^y,
	 \; ZZ=(i/2)\s_1^z\s_2^z,
	 \; YZ=(i/2)\s_1^y\s_2^z,
	 \; ZY=(i/2)\s_1^z\s_2^y,
\EEqA
with $k=1,2$. The operators listed above form the Lie algebra 
$L_0 = {\rm span}
\left\{X_1, X_2, XX, YY, ZZ, YZ, ZY \right\}\subset {\rm su(4)}$ whose 
commutators are given in the following table:
\begin{equation}
\label{eq:COMMTABLEk}
 \begin{array}{c|ccccccc}\hline
    & X_1 & X_2 & XX& YY & ZZ &  YZ		&		ZY \\ \hline
X_1 &  0  &  0  & 0 & -ZY& YZ &  -ZZ	&		YY  \\
X_2 &  0  &  0  & 0 & -YZ& ZY & 	YY	&	-ZZ\\ 
XX  &  0  &  0  & 0 &  0 & 0  &    0	&	0\\ 
YY &    ZY&  YZ & 0 &  0 & 0  &		-X_2&-X_1\\ 
ZZ &  -YZ  & -ZY& 0 &  0 & 0  & 	X_1	&X_2\\ 
YZ &  ZZ   & -YY& 0 & X_2&-X_1&  		0	&	0 \\ 
ZY & -YY   &  ZZ& 0 & X_1&-X_2&  		0	&	0 \\ \hline
 \end{array}
\end{equation}
Later on, in sections devoted to applications, we will revert to the 
usual notation.

Notice that it is possible to generate Lie algebras isomorphic to 
(\ref{eq:COMMTABLEk}) by replacing the local operators $(X_1, X_2)$ 
with either $(Y_1, Y_2)$ or $(Z_1, Z_2)$, without any change in our 
results. An example of this will be given in Sec. \ref{sec:Z1mZ2}. 

\section{General considerations}

Let us consider a generic time-dependent Hamiltonian
\BEqA
iH(t) = 
\Omega_{1x}(t)X_1+\Omega_{2x}(t)X_2
+g_{xx}(t)XX
+g_{yy}(t)YY+g_{zz}(t)ZZ 
+ g_{yz}(t)YZ + g_{zy}(t)ZY,
\EEqA
whose scalar functions will be called the {\it steering controls}, 
or {\it control parameters}. 
The solution to the Schr{\"o}dinger equation
\BEq
\label{eq:SCHRODEQ}
\frac{dU(t)}{dt} = -iH(t)U(t),\quad U(0)=1,
\EEq
is a time-dependent operator $U(t)\in \exp(L_0)\subset {\rm SU(4)}$,
which can always be written in the form \cite{MYCNOTPAPER1}
\BEqA
\label{eq:UCARTAN}
U(t)= \underbrace{e^{-\a(t)X_1-\b(t)X_2}}_{k_1(t)}
	\underbrace{e^{-c_1(t)XX-c_2(t)YY-c_3(t)ZZ}}_{U_{\rm ent}(t)}
 \underbrace{e^{-\zeta(t)X_1-\xi(t)X_2}}_{k_2(t)}.
\EEqA
The functions appearing
in the exponents of Eq. (\ref{eq:UCARTAN}) will be collectively 
referred to as the {\it steering parameters}, while the triplet 
$(c_1(t), c_2(t), c_3(t))$ will be called the {\it class vector}, 
as usual \cite{TUCCI}. In what follows, the class vector will be 
allowed to evolve on the full Cartan subalgebra 
$A_C = {\rm span}\left\{XX, YY, ZZ\right\}\subset L_0 \subset 
{\rm su(4)}$ rather than within the Weyl chamber, since 
projecting it onto the Weyl chamber can always be easily performed 
\cite{TUCCI}. It is important to remember that at any 
given time $t$ the choice of the steering parameters is 
{\it not} unique. Therefore, additional requirements (such as 
smoothness, initial conditions, etc.) must be imposed on the 
corresponding functions in order to determine the experimentally 
meaningful trajectory.

Differentiating (\ref{eq:UCARTAN}) with respect to the time $t$ 
gives:
\BEqA
\label{eq:DUDT}
\frac{dU(t)}{dt} &=& -[\a'X_1 + \b'X_2 + c_1' XX+ 
					e^{-\a X_1}e^{-\b X_2}\left(c_2'YY+c_3'ZZ\right)
										e^{\b X_2}e^{\a X_1} \nonumber \\
&& \quad \quad + 	e^{-\a X_1}e^{-\b X_2}e^{-c_2YY}e^{-c_3ZZ}
					\left(\zeta'X_1+\xi'X_2\right)
										e^{c_3ZZ}e^{c_2YY}e^{\b X_2}e^{\a X_1}
																				]\; U(t).					
\EEqA
Here, each of the nested similarity transformations represents 
a rotation by some angle in a certain two-dimensional subspace 
of the Lie algebra $L_0$.
For instance,
\BEqA
	\label{eq:ROTATIONS}
	e^{-c_3 ZZ}X_1 e^{c_3 ZZ} &=& X_1 \cos c_3  + YZ \sin c_3 ,
	\nonumber \\
	e^{-c_2 YY}YZ e^{c_2 YY} &=& YZ \cos c_2 + X_2 \sin c_2,
	\nonumber \\
	e^{-\a X_1}YZ e^{\a X_1} &=& YZ \cos \a  + ZZ \sin \a ,
\EEqA
etc. Using (\ref{eq:ROTATIONS}) to perform algebraic 
manipulations in (\ref{eq:DUDT}) and equating the resulting 
coefficients of the corresponding generators on the right hand 
sides of (\ref{eq:SCHRODEQ}) and (\ref{eq:DUDT}), we get a 
nonlinear system of seven first-order differential equations
\BEqA
\label{eq:MAINSYSTEM1}
\underbrace{
\left[
\begin{array}{ccccccc}
1 & 0 & 0 & 0 & 0 & 0   & 0	\\
0 & 1 & 0 & 0 & 0 & C_1 & C_2  	\\
0 & 0 & 1 & 0 & 0 & C_2 & C_1 	\\
0 & 0 & 0 & A_1 & A_2 & -A_3C_3+A_4C_4 & A_3C_4-A_4C_3 \\
0 & 0 & 0 & A_2 & A_1 & A_4C_3-A_3C_4 & -A_4C_4+A_3C_3 \\
0 & 0 & 0 & A_3 & -A_4 & A_1C_3+A_2C_4 & -A_1C_4-A_2C_3 \\
0 & 0 & 0 & A_4 & -A_3 & -A_2C_3-A_1C_4 & A_2C_4+A_1C_3 \\
\end{array}
\right]
}_{M}
\left[
\begin{array}{c}
c_1' \\
\a'	\\
\b'	\\
c_2' \\
c_3'\\
\zeta'\\
\xi'\\
\end{array}
\right]=
\left[
\begin{array}{c}
g_{xx} \\
\Omega_{1x}\\
\Omega_{2x}\\
g_{yy}\\
g_{zz}\\
g_{yz}\\
g_{zy}\\
\end{array}
\right],
\EEqA
where the new variables
\BEqA
	C_1 =  \cos c_2 \cos c_3, \quad C_2 = \sin c_2 \sin c_3, \quad
	C_3 = \cos c_2 \sin c_3, \quad	C_4 = \sin c_2 \cos c_3,
\EEqA
and
\BEqA
		A_1 = \cos \a \cos \b, \quad
		A_2 = \sin \a \sin \b, \quad
		A_3 = \cos \a \sin \b, \quad
		A_4 = \sin \a \cos \b,
\EEqA
have been introduced. Notice that
\BEq
\label{eq:detM}
\det{M} = \cos^2 c_2-\cos^2 c_3.
\EEq
For simplicity, we choose
\BEqA
\label{eq:incond}
c_1(0) = \a(0) = \b(0) = c_2(0) = c_3(0) 
= \zeta(0) = \xi(0) = 0
\EEqA
to satisfy the initial condition $U(0)=1$. 

The first equation in (\ref{eq:MAINSYSTEM1}) integrates immediately,
\BEq
c_1(t)=\int_0^t d\t \; g_{xx}(\t),
\EEq
while the remaining system can be inverted to give
\BEq
\label{eq:MAINSYSTEM2}
\left[
\begin{array}{c}
\a'	\\
\b'	\\
c_2' \\
c_3'\\
\zeta'\\
\xi'\\
\end{array}
\right]=
\left[
\begin{array}{c}
\Omega_{1x}
+\frac{
g_{yy}(A_3C_{33}+A_4C_{22})
-g_{zz}(A_3C_{22}+A_4C_{33})
-g_{yz}(A_1C_{33}-A_2C_{22})
-g_{zy}(A_1C_{22}-A_2C_{33})
}{\det{M}}
\\
\Omega_{2x}
+\frac{
g_{yy}(A_3C_{22}+A_4C_{33})
-g_{zz}(A_3C_{33}+A_4C_{22})
-g_{yz}(A_1C_{22}-A_2C_{33})
-g_{zy}(A_1C_{33}-A_2C_{22})
}{\det{M}}
\\
 g_{yy} A_1 +  g_{zz} A_2 +  g_{yz} A_3 +  g_{zy} A_4\\
g_{yy} A_2+  g_{zz} A_1 -  g_{yz} A_4 -  g_{zy} A_3\\
\frac{-g_{yy}(A_3 C_3 + A_4 C_4)
 + g_{zz} (A_3 C_4 + A_4 C_3)
 + g_{yz} (A_1 C_3 - A_2 C_4)
 + g_{zy} (A_1 C_4 - A_2 C_3)
}{\det{M}}
\\
\frac{-g_{yy}  (A_3 C_4 + A_4 C_3)
 + g_{zz} (A_3 C_3 + A_4 C_4)
 + g_{yz} (A_1 C_4 - A_2 C_3)
 + g_{zy} (A_1 C_3 - A_2 C_4)
}{\det{M}}
\\
\end{array}
\right],
\EEq
where
\BEqA
	C_{22} =  \cos c_2\sin c_2, \quad 
			 C_{33} = \cos c_3 \sin c_3.
\EEqA
To make further progress, we impose some 
restrictions on the form of the steering Hamiltonian.

\section{Anisotropic exchange with tracking controls}

{\it Tracking} \cite{MYCNOTPAPER2} refers to steering with control 
parameters having
the same enveloping profile defined by some function $\g(t)$.
Notice that any time-independent Hamiltonian describes tracking 
with $\g(t)=1$. Here we are interested in Hamiltonians 
\BEqA
	\label{eq:ansatz hamiltonian}
	iH(t) = \g(t)[\Omega_1X_1+\Omega_2X_2	+g_1(t)XX
		+g_2YY+g_3ZZ],
\EEqA
where $g_1(t)$ is a function of time and
$\Omega_{1}, \Omega_{2}, g_2, g_3$ are some constants. 
[It is possible to choose $g_1(t)$ arbitrarily because $XX$ is 
central in $L_0$.]

\subsection{Solving the tracking control case}

Under these conditions,
\BEq
\label{eq:trackingc1}
c_1(t)=\int_0^t d\t \; \g(\t)g_1(\t).
\EEq
The remaining steering parameters will be found using the
ansatz
\BEq
\label{eq:ansatz}
\a(t)=\zeta(t),\; \b(t)=\xi(t),
\EEq
or, equivalently,
\BEqA
\label{eq:UCARTANansatz}
U(t)= e^{-\a(t)X_1-\b(t)X_2}e^{-c_1(t)XX-c_2(t)YY-c_3(t)ZZ}
 e^{-\a(t)X_1-\b(t)X_2}.
\EEqA
This ansatz works only for Hamiltonians given in 
(\ref{eq:ansatz hamiltonian}). For more general systems, another 
trick or numerical simulations based on (\ref{eq:UCARTAN}) and 
(\ref{eq:MAINSYSTEM2}) should be used. 

The resulting system is
\BEqA
\left[
\begin{array}{c}
\a'	\\
\b'	\\
c_2' \\
c_3'\\
\a'\\
\b'\\
\end{array}
\right]=
\g
\left[
\begin{array}{c}
\Omega_{1}
+\frac{
(g_{2}A_4
-g_{3}A_3)C_{22}
+(g_{2}A_3-g_{3}A_4)C_{33}
}
{\det{M}}
\\
\Omega_{2}+
\frac{
(g_{2}A_3
-g_{3}A_4)C_{22}
+(g_{2}A_4
-g_{3}A_3)C_{33}
}{
\det{M}
}
\\
 g_{2} A_1 +  g_{3} A_2 \\
g_{2} A_2+  g_{3} A_1 \\
 \frac{
-(g_{2}A_4-g_{3}A_3)C_4
-(g_{2}A_3-g_{3}A_4)C_3
}
{\det{M}}
\\
 \frac{
-(g_{2}A_4-g_{3}A_3)C_3
-(g_{2}A_3-g_{3}A_4)C_4 
}
{\det{M}}\\
\end{array}
\right].
\EEqA
The four equations for $\a'$ and $\b'$ give
	\BEqA
	\label{eq:ABPRIME}
	\a'= \frac{\g \left[\Omega_1(1+\cos c_2 \cos c_3)
	-\Omega_2\sin c_2 \sin c_3\right]}
					{(\cos c_2 + \cos c_3)^2}, 
					\quad
	\b'=\frac{\g \left[\Omega_2(1+\cos c_2 \cos c_3)
	-\Omega_1\sin c_2 \sin c_3\right]}
					{(\cos c_2 + \cos c_3)^2},
	\EEqA
which determine $\a(t)$ and $\b(t)$ after
$c_2(t)$ and $c_3(t)$ had been found. Also,
\BEqA
			A_3 =  \frac{
		(\Omega_1 g_3- \Omega_2 g_2 ) \sin c_2 
		- (\Omega_1 g_2-\Omega_2 g_3 ) \sin c_3 }
		 {(g_2^2  - g_3^2 ) (\cos c_2  + \cos c_3 )}, 
		 \quad
		A_4 =  \frac{
		(\Omega_1 g_3 - \Omega_2 g_2) \sin c_3
		-	(\Omega_1 g_2 - \Omega_2 g_3) \sin c_2  }
		 {(g_2^2  - g_3^2 ) (\cos c_2  + \cos c_3 )}.
\EEqA
The equations for $c_2'$ and $c_3'$ give
\BEq
\left(c_2'\pm c_3'\right)^2
=
 \g^2 (g_2\pm g_3)^2(A_1 \pm A_2)^2.
\EEq
Using
\BEq
\label{eq:Aeq}
(A_1 \pm A_2)^2 = 1 - (A_3 \mp A_4)^2,
\EEq
we get
\BEqA
\label{eq:c2pmc3sq}
\left(c_2'\pm c_3'\right)^2
&=& \g^2 \left[
(g_2\pm g_3)^2
 - \frac{\left(\Omega_1\mp \Omega_2\right)^2\left(\sin c_2 
	\pm \sin c_3\right)^2}{(\cos c_2 + \cos c_3)^2}
	\right].
\EEqA
After applying the sum-to-product identities and re-arranging the 
terms, we arrive at
\BEqA
\label{eq:mainequation}
\left(\frac{d}{dt}\sin \left(\frac{c_2(t) \pm c_3(t)}{2}
\right)\right)^2 
+ \left( \frac{\g(t)}{2}
\sqrt{(g_2\pm g_3)^2+\left(\Omega_1 \mp \Omega_2\right)^2}
	\sin \left(\frac{c_2(t) \pm c_3(t)}{2}\right)\right)^2
	= \left(\frac{\g(t)}{2}(g_2\pm g_3)\right)^2.
\EEqA
By making substitution
\BEq
f_{\pm}(t) := \sin \left(\frac{c_2(t) \pm c_3(t)}{2}\right),
\EEq
we can solve the resulting equation
\BEqA
	\label{eq:substituteseq}
	\left(\frac{df_{\pm}(t)}{dt}\right)^2 
	+ \left(\frac{\g(t)}{2}
	\sqrt{(g_2\pm g_3)^2+\left(\Omega_1 \mp \Omega_2\right)^2}
		f_{\pm}(t)\right)^2
		= \left(\frac{\g(t)}{2}(g_2\pm g_3)\right)^2
\EEqA
by inspection. It is easy to see that
\BEqA
	f_{\pm}(t)&=&
	\frac{g_2\pm g_3}{\sqrt{\left(g_2\pm g_3\right)^2+
	\left(\Omega_1 \mp \Omega_2\right)^2}}
 \sin\left( 
	\frac{\sqrt{(g_2\pm g_3)^2+\left(\Omega_1 \mp \Omega_2\right)^2}}{2}
	\int_0^t \g(\t)d\t
	\right)
\EEqA
solves (\ref{eq:substituteseq}) subject to (\ref{eq:incond}), which 
together with 
(\ref{eq:trackingc1}), (\ref{eq:UCARTANansatz}),
(\ref{eq:ABPRIME}) solves the tracking control case:
\BEq
\label{eq:solution}
c_{2,3}(t) = \arcsin\left(f_{+}(t)\right)\pm \arcsin
\left(f_{-}(t)\right).
\EEq 

\subsection{Controlling the flow on the Weyl chamber}

Let us now assume that $g_1$ is tunable, but otherwise independent of 
time.  Then, given an experimentally realizable tracking mechanism $\g(t)$, 
a point on the Weyl chamber (alternatively, in the full Cartan subalgebra 
$A_C$) whose $XX$ coordinate is $c_1$ can be reached after steering for 
a time $t_1$ satisfying
\BEq
\int_0^{t_{1}} \g(\t) d\t = \frac{c_1}{g_1}.
\EEq
This reachability condition is necessary, but not sufficient. Since the 
point is specified by a class vector $\vec{c}=(c_1,c_2,c_3)$, we have 
yet to determine whether the remaining coordinates $c_{2,3}$ can be 
realized by adjusting the Rabi frequencies $\Omega_{1,2}$. 
Here we will restrict our attention only to the points belonging to the 
$XX$ axis, and thus having $c_{2,3}=0$.
It is easy to see that in this case we must have
\BEq
\Omega_{1,2}^{(c_1, 0, 0)} = \frac{1}{2}\left[
\sqrt{\left(\frac{2\pi n g_1}{c_1}\right)^2-(g_2-g_3)^2}
\pm\sqrt{\left(\frac{2\pi m g_1}{c_1}\right)^2-(g_2+g_3)^2}\right],
\EEq
provided the integers $n$, $m$ are chosen in such a way as to make the 
Rabi frequencies real. 

We can now write down a general condition under which a coupled qubit 
system directly generates controlled-NOT class corresponding to 
$\vec{c}=\pi/2\times(1, 0, 0)$:
\BEqA
\int_0^{t_{\rm CNOT}} \g(\t)d\t = \frac{\pi}{2g_1}, 
\quad \Omega_{1,2}^{\rm CNOT} = \frac{1}{2}
\left[\sqrt{(4ng_1)^2-(g_2-g_3)^2}
\pm\sqrt{(4mg_1)^2-(g_2+g_3)^2}\right].
\EEqA
Other approaches to CNOT gate design have been considered in Refs. 
\cite{WELLCNOT, RIGETTI2005, GRIGORENKO2005, HILL2007}.

\subsection{Tracking control of Josephson phase qubits}
  
\subsubsection{Capacitive coupling with rf bias of $\Omega_1\s_1^x$ 
type}

In the rotating wave approximation (RWA) \cite{REDFIELD1955} the dynamics of two resonant
capacitively 
coupled phase qubits
\cite{JOHNSONETAL2003, BLAIS2003, WELLCNOT, 
MARTINIS_Science2005, MYTUNNELINGPAPER, KOFMAN_ZHANG_MARTINIS_2007} 
is described by the Hamiltonian
\BEqA
\label{eq:capacitive coupling}
H_1(t) = (\g(t)/2)
[\Omega_1\s_1^x+g(\s_1^x\s_2^x+\s_1^y\s_2^y)], \quad g> 0.
\EEqA
The Rabi term represents the action of an rf bias current applied to 
one of the qubits. It turns out that keeping just one such local 
term suffices to generate controlled-NOT 
logic \cite{MYCNOTPAPER1}. The condition 
$\Omega_2=g_3=0$ gives $f_{+}(t)=f_{-}(t)$, which leads to
\BEqA
c_1(t) = g \int_0^t \g(\t)d\t,
\quad
c_2(t) =
2\arcsin
\left[
\frac{1}{\sqrt{1+(\Omega_1/g)^2}}
\sin\left( \frac{g}{2}\sqrt{1+(\Omega_1/g)^2}
\int_0^t \g(\t)d\t
\right)
\right], \quad
c_3(t) = 0, 
\EEqA
and
\BEqA
 \a(t) = \Omega_1 \int_0^t d\t\; \frac{\g}{1 + \cos c_2}, 
\quad \b(t)=0.
\EEqA
The time-dependent gate is therefore
\BEqA
	U(t)= e^{-(i/2)\a\s_1^x}
	e^{-(i/2)\left(c_1\s_1^x\s_2^x+c_2\s_1^y\s_2^y\right)}e^{-(i/2)\a\s_1^x},
\EEqA
which becomes an element of controlled-NOT class, provided 
\cite{MYCNOTPAPER1}
\BEqA
\int_0^{t_{\rm CNOT}} \g(\t)d\t = \pi/2g,
\quad
\Omega_1^{\rm CNOT} = g\sqrt{(4n)^2-1},
\EEqA
with $n = 1,2,3,\dots$.

We may use Result 1 of Ref. \cite{ANGELO2005} to state the following 
applicability condition for the RWA: \\
The solution to the Schr{\"o}dinger equation with the RWA Hamiltonian 
(\ref{eq:capacitive coupling}) approximates the solution with exact 
$H$ (reduced to computational subspace; see Ref. \cite{MYCNOTPAPER1} 
for details) in the sense that if $\Omega_1/\omega \ll 1$
(weak perturbation) and $\omega = \epsilon$ (resonant condition), then 
$||\psi_{\rm RWA}(t)- \psi_{\rm exact}(t)||= O(\Omega_1/\omega)$
whenever
$0\leq t \leq O(\omega/\Omega_1)$. Here, $\omega$ is the bias 
frequency, $\epsilon$ is the computational level splitting.
For UCSB architectures \cite{MARTINIS_GELLER_private_comm} with qubit 
coupling $g\sim 10$ GHz and level splitting $\omega \lesssim 100$ MHz, 
$\Omega_1^{\rm CNOT}/\omega \sim 10^{-2}$.

For calculations that go beyond the RWA in the context of Josephson phase 
qubits coupled to nanomechanical resonators see Ref. \cite{GELLER2004}.

\subsubsection{Inductive coupling with rf bias of $\Omega_1\s_1^x+
\Omega_2\s_2^x$ type}

For inductively coupled qubits 
\cite{PLOURDE, MAJER_MOOIJ_2005, MAASSEN_BRINK_2005, GRANATA2005, 
SIDDIQI_CLARKE_2006,
LIU_WEI_TSAI_NORI_2006, YOU_NAKAMURA_NORI_2005, ZHOU_CHU_HAN_2006} 
driven by local rf magnetic fluxes the 
Hamiltonian in the RWA is \cite{MYCNOTPAPER1}
\BEqA
	\label{eq:inductive coupling}
	H_2(t) = (\g(t)/2)
	[\Omega_1\s_1^x+\Omega_2\s_2^x + g(\s_1^x\s_2^x
		+\s_1^y\s_2^y
			+k\s_1^z\s_2^z)], \quad g>0.
\EEqA
Using (\ref{eq:solution}), the steering trajectory is found to be
\BEqA
c_1(t) &=& g \int_0^t \g(\t)d\t, \nonumber \\
c_{2,3}(t)&=&
\arcsin 
\left[
\frac{1+k}{\sqrt{(1+k)^2+[(\Omega_1 - \Omega_2)/g]^2}}
\sin\left( \frac{g}{2}
\sqrt{(1+k)^2+[(\Omega_1 - \Omega_2)/g]^2} 
\int_0^t \g(\t)d\t
\right)
\right] \nonumber \\
&& \quad \pm
\arcsin 
\left[
\frac{1-k}{\sqrt{(1-k)^2+[(\Omega_1 + \Omega_2)/g]^2}}
\sin\left( \frac{g}{2}
\sqrt{(1-k)^2+[(\Omega_1 + \Omega_2)/g]^2} 
\int_0^t \g(\t)d\t
\right)
\right],
\EEqA
where
\BEqA
	U(t)
	= e^{-(i/2)(\a\s_1^x+\b\s_2^x)}
	e^{-(i/2)\left(c_1\s_1^x\s_2^x+c_2\s_1^y\s_2^y+c_3\s_1^z\s_2^z\right)}
	e^{-(i/2)(\a\s_1^x+\b\s_2^x)},
\EEqA
with $\a$ and $\b$ calculated from Eq. (\ref{eq:ABPRIME}).
The CNOT class is generated by setting 
\cite{MYCNOTPAPER2}
\BEqA
\int_0^{t_{\rm CNOT}} \g(\t)d\t = \frac{\pi}{2g}, 
\quad 
\Omega_{1,2}^{\rm CNOT}
= \frac{g}{2}\left[\sqrt{(4n)^2-(1-k)^2} \pm\sqrt{(4m)^2-(1+k)^2}\right].
\EEqA

\begin{figure}
\includegraphics{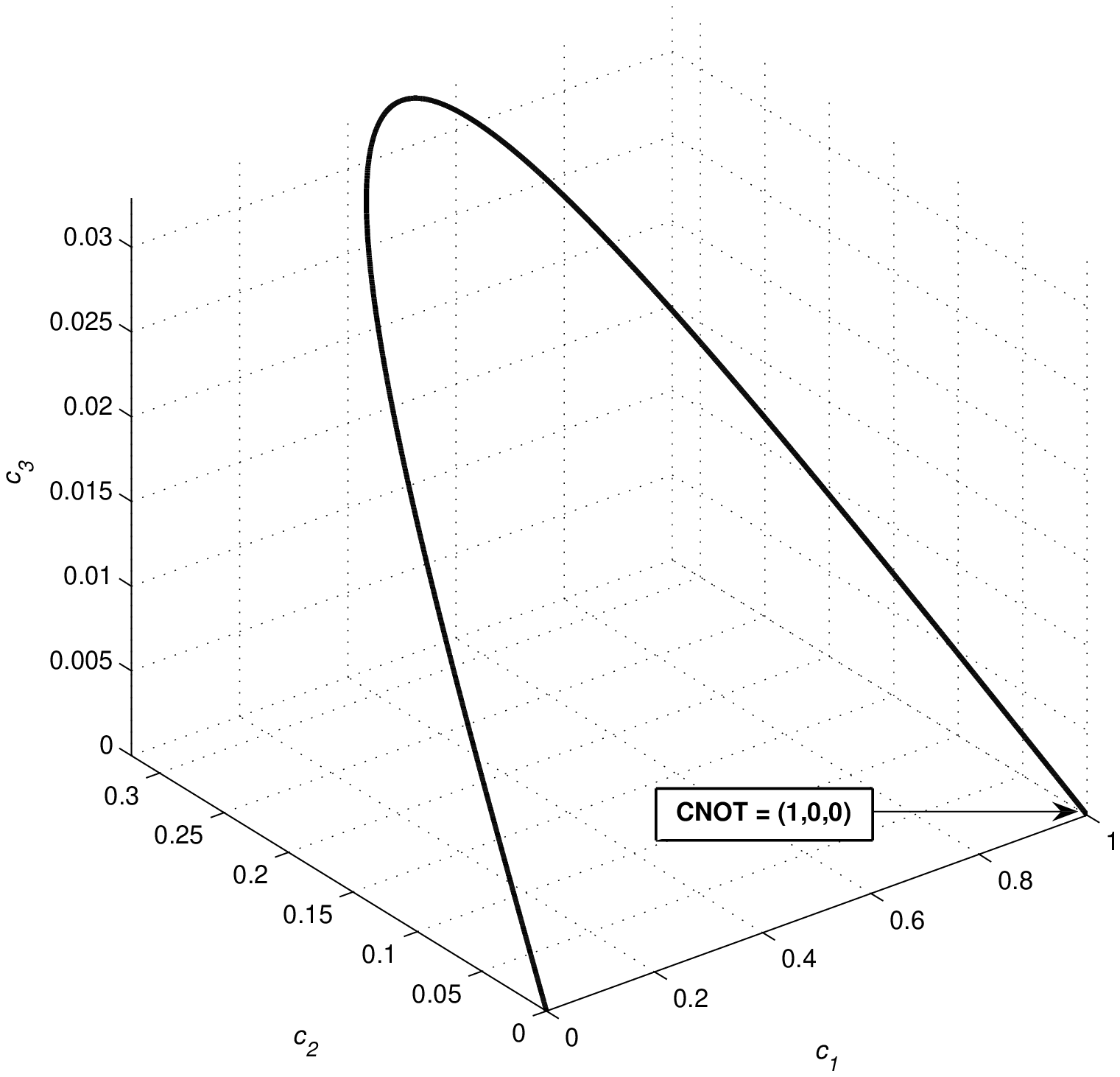}
\caption{
\label{fig:1} 
Steering trajectory generating CNOT class
in the case of rf-biased inductively coupled
flux qubits, Eq. (\ref{eq:inductive coupling}). Here, $g=1.00$, 
$k=0.10$, $\Omega_1 = 3.8716$, $\Omega_2 = 
0.0258$, $\g(t)=1$. The steering parameters are given in units of 
$\pi/2$.
}
\end{figure}

\begin{figure}
\includegraphics{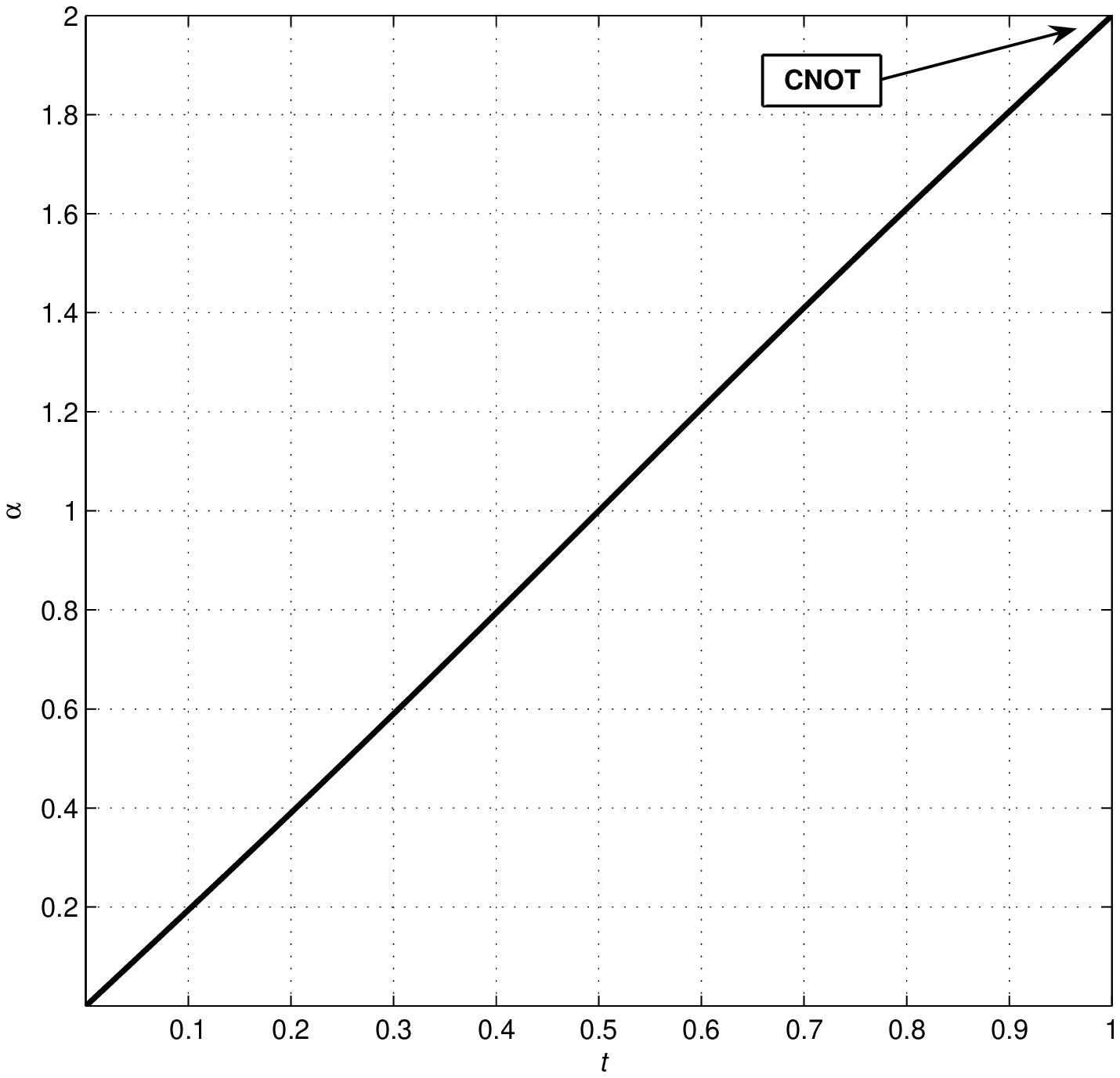}
\caption{
\label{fig:2} 
Local rotation $\alpha$ accompanying the steering 
trajectory 
shown in Fig. \ref{fig:1}.
}
\end{figure}

\begin{figure}
\includegraphics{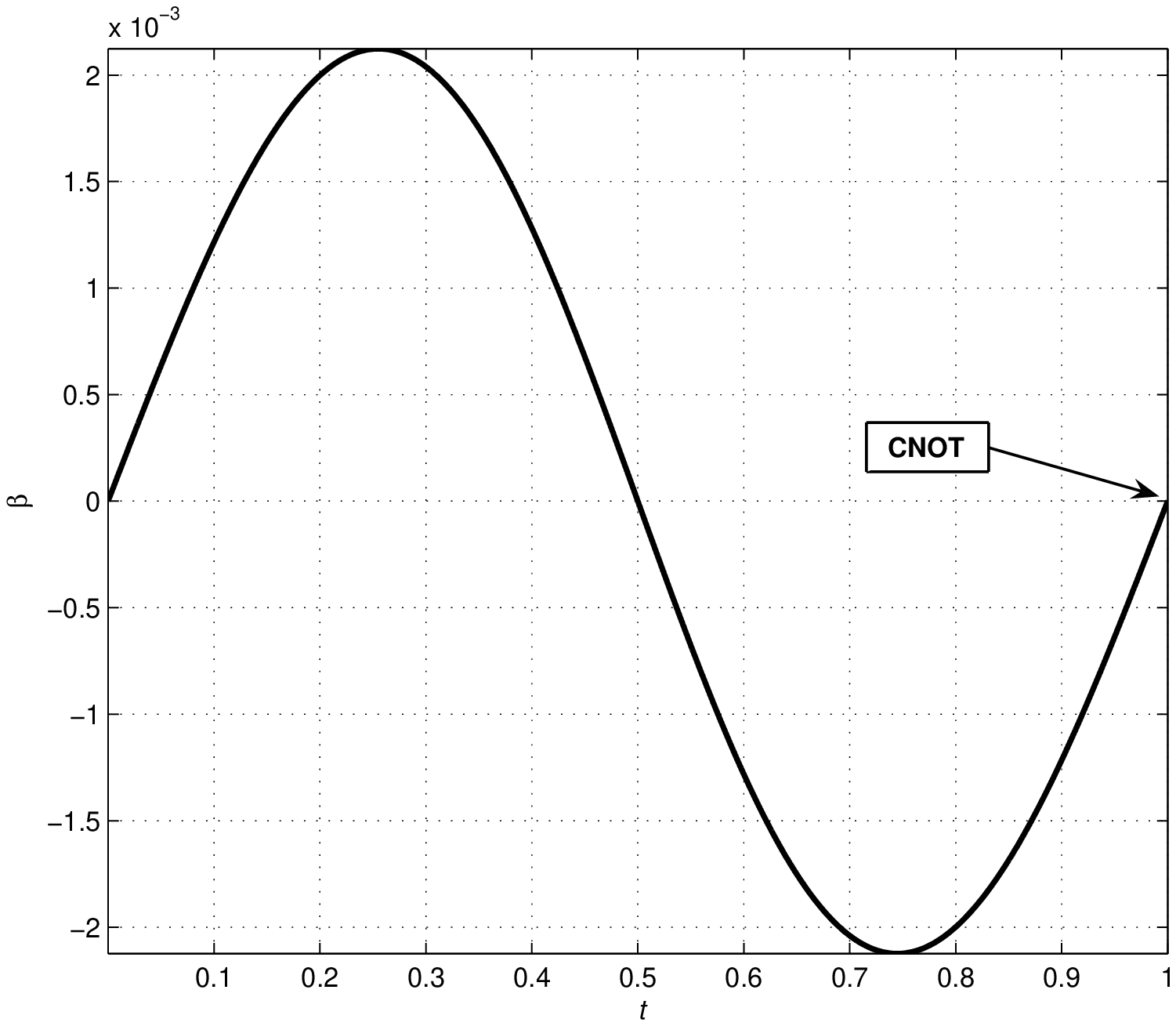}
\caption{
\label{fig:3} 
Local rotation $\beta$ accompanying the steering 
trajectory 
shown in Fig. \ref{fig:1}.
}
\end{figure}

For example, for $g=1.00$, $k=0.10$ and $n=m=1$, the Rabi frequencies are 
$\Omega_1 = 3.8716$, $\Omega_2 = 0.0258$. The corresponding Weyl chamber 
steering trajectory for $\g(t)=1$, with parameters measured in units of 
$\pi/2$, is shown in Figures \ref{fig:1}, \ref{fig:2}, and \ref{fig:3}.

\subsubsection{Inductive coupling with dc bias of $\Omega_1(\s_1^z-\s_2^z)$ 
type}
\label{sec:Z1mZ2}

Because of the $(X_1,X_2)\rightarrow (Y_1,Y_2) \rightarrow (Z_1,Z_2)$ 
``symmetry'' mentioned in Sec. \ref{sec:INTRO}, it is possible to devise 
an alternative CNOT implementation based on the Hamiltonian for inductively 
coupled qubits acted upon by dc fluxes:
\BEqA
	\label{eq:inductive coupling alt}
	H_3(t) = (\g(t)/2)
	[\Omega_1(\s_1^z-\s_2^z)
 			+g(k\s_1^z\s_2^z 
 			 + \s_1^x\s_2^x + \s_1^y\s_2^y)].
\EEqA
The effect of such bias is to ``move'' system's energy levels by equal 
amounts in opposite directions (the process known as {\it detuning}).
One important feature of this implementation is that for any $|k|<1/2$ 
it is always possible to generate controlled-NOT logic by choosing Rabi 
frequencies $0<|\Omega_1|/g<1$. This is important when
perturbation is required to be small 
(see Section \ref{sec:REDUCING_LEAKAGE} for a more general approach). 

We have,
\BEqA
c_1(t) = kg \int_0^t \g(\t)d\t, \quad
c_2(t) = c_3(t) =
\arcsin
\left[
\frac{1}{\sqrt{1+(\Omega_1/g)^2}}
\sin\left( g\sqrt{1+(\Omega_1/g)^2}
\int_0^t \g(\t)d\t
\right)
\right], 
\EEqA
and
\BEqA
 \a(t) = - \b(t)=\frac{\Omega_1}{2} \int_0^t d\t \;
\frac{\g}{\cos^2 c_2} \;,
\EEqA
where the steering parameters $(\a, \b, c_1, c_2, c_3)$ are now 
associated with the operators $(Z_1, Z_2, ZZ, XX, YY)$.
The time-dependent gate is given by
\BEqA
	U(t)= e^{-(i/2)\a(\s_1^z-\s_2^z)}
	e^{-(i/2)\left(c_1\s_1^z\s_2^z+c_2(\s_1^x\s_2^x+\s_1^y\s_2^y)\right)}
	 e^{-(i/2)\a(\s_1^z-\s_2^z)},
\EEqA
which implements CNOT class, provided
\BEq
\int_0^{t_{\rm CNOT}} \g(\t)d\t = \pi/(2kg), 
\quad \Omega_{1}^{\rm CNOT} = g\sqrt{(2kn)^2-1},
\EEq
where $(2kn)^2>1$.
Several examples of this implementation are listed in
Table \ref{tab:table1}.

\begin{table}
\caption{
\label{tab:table1} 
Generation of controlled-NOT logic 
with $\vec{c} = (\pi/2)\times (1,0,0)$
using inductively coupled flux qubits subject to symmetric dc detuning 
$(\Omega_1/2)(\s_1^z-\s_2^z)$. The Hamiltonian is given in 
(\ref{eq:inductive coupling alt}). Here, $\g(t)=1$.
}
\begin{ruledtabular}
\begin{tabular}{lccc}
$k$ & $t_{\rm CNOT}$, units of $\pi/2g$ & $n$ & $\Omega_1^{\rm CNOT}/g$  
\\ \hline
 0.100 	& 10	& 6 	& 0.6633 				\\ 
  			& 		&7 		& 0.9798			 	\\
	 	  	& 		&8 		& 1.2490				\\ \hline
 0.050	& 20	&11 	& 0.4583	 			\\ 
				& 		&12 	& 0.6633				\\
				& 		&13 	& 0.8307				\\ 
				& 		&14 	& 0.9798				\\
				& 		&15 	& 1.1180				\\ \hline
 0.025	& 40	&21 	& 0.3202	 			\\ 
				& 		&22 	& 0.4583	 			\\
				& 		&23 	& 0.5679				\\ 
				& 		&24 	& 0.6633				\\
				& 		&25 	& 0.7500				\\ 
				& 		&26 	& 0.8307				\\ 
				& 		&27 	& 0.9069				\\
				& 		&28 	& 0.9798				\\ 
				& 		&29 	& 1.0500				\\ \end{tabular}
\end{ruledtabular}
\end{table}

\begin{figure}
\includegraphics{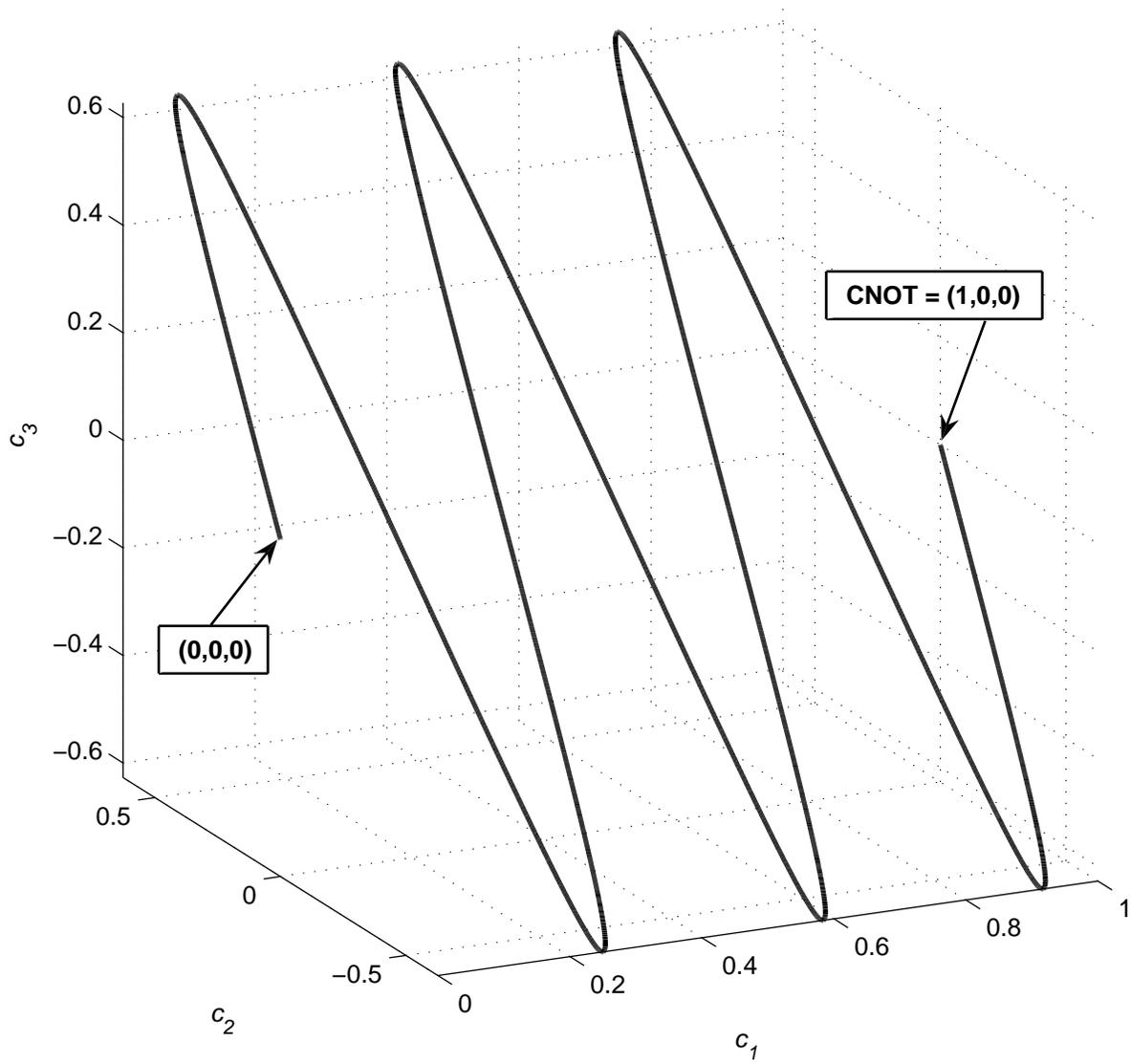}
\caption{
\label{fig:4} 
Steering trajectory generating CNOT 
class in the case of inductively coupled
flux qubits subject to dc symmetric detuning, 
Eq. (\ref{eq:inductive coupling alt}). Here, $g=1.00$, 
$k=0.10$, $\Omega_1 = 0.6633$, $\g(t)=1$. 
The steering parameters are given in units of $\pi/2$.
}
\end{figure}

\begin{figure}
\includegraphics{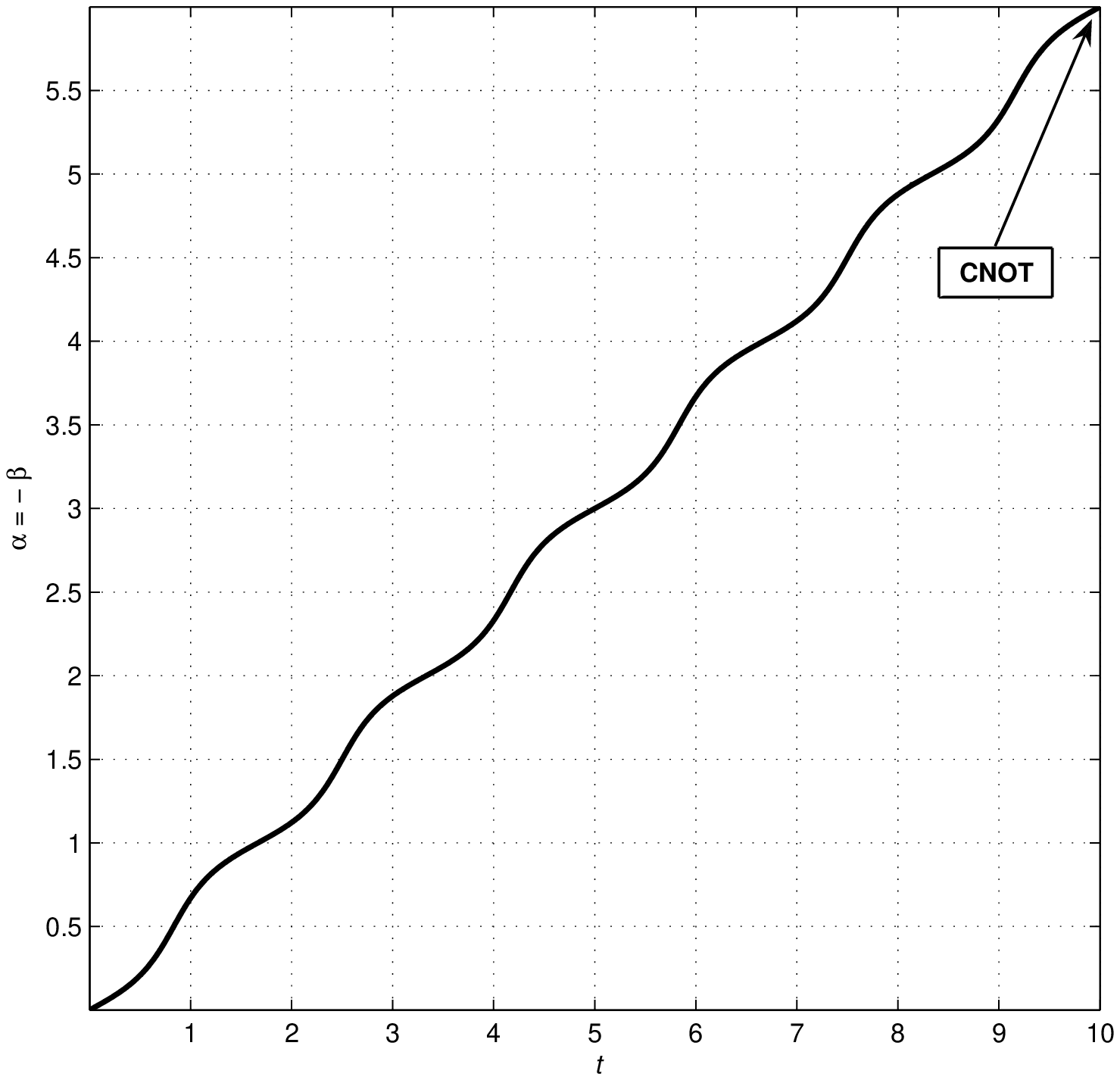}
\caption{
\label{fig:5} 
Local rotations accompanying the steering 
trajectory shown in Fig. \ref{fig:4}.
}
\end{figure}

Figures \ref{fig:4}, \ref{fig:5} show the steering trajectory for 
$g=1.00$, $k=0.10$, and the Rabi frequency $\Omega_1 = 0.6633$.

\section{Discussion}

We now discuss limitations and possible 
extensions of the proposed method.

The most significant limitation comes from restricting the local 
terms to form a {\it homogeneous} pair (such as, for example, 
$(X_1, X_2)$).
By adopting such restriction we were able to isolate a special 
subalgebra $L_0$ of ${\rm su(4)}$, given in Eq. (\ref{eq:COMMTABLEk}),
that contains a central element.
The fifteen-dimensional problem was then reduced to a nonlinear system 
of 
``only'' seven first-order differential equations, one of which 
completely separated from the others. By making a certain 
ansatz, the analytical solution in the tracking 
control case has been found.

We can extend this approach to Hamiltonians with arbitrary 
combinations of Rabi terms, such as $(X_1, Y_2)$, etc. 
The dimensionality of the problem would increase, but it would still 
be possible to write down and solve --- most likely, numerically --- 
the corresponding 
system of differential equations.

For 
Hamiltonians containing {\it homogeneous} local terms {\it with arbitrary time 
dependence} the following useful ansatz can be identified:

{\it Case 1.} For
\BEq
	\label{eq:beyond tracking hamiltonian}
	H(t) = (1/2)[\Omega_1(t)\s_1^x+g_1(t)\s_1^x\s_2^x+g_2(t)\s_1^y\s_2^y],
\EEq
use
\BEq
\label{eq:ansatz beyond 1}
c_3(t)=0, \quad \b(t)=\xi(t)=0,
\EEq
which corresponds to the Cartan decomposition
\BEqA
	\label{eq:UCARTANansatz beyond 1}
	U(t)= e^{-(i/2)\a\s_1^x}e^{-(i/2)(c_1\s_1^x\s_2^x+c_2\s_1^y\s_2^y)}
	e^{-(i/2)\zeta \s_1^x}.
\EEqA
Eq. (\ref{eq:MAINSYSTEM2}) then reduces to
\BEqA
\label{eq:MAINSYSTEM3}
\left[
\begin{array}{c}
\a'	\\
c_2' \\
\zeta'\\
\end{array}
\right]=
\left[
\begin{array}{c}
\Omega_{1}
-g_{2}\sin \a \cos c_2/\sin c_2
\\
 g_{2} \cos \a \\
g_{2}\sin \a/\sin c_2 
\\
\end{array}
\right].
\EEqA

{\it Case 2. Anisotropic exchange with symmetric detuning.}
This case generalizes the detuning Hamiltonian considered in Section 
\ref{sec:Z1mZ2} by allowing arbitrary time dependent controls: 
\BEqA
	\label{eq:inductive coupling alt general}
	H(t) = (1/2)[\Omega_1(t)(\s_1^z-\s_2^z)
	 +g_1(t)\s_1^z\s_2^z 
	  + g_2(t)\s_1^x\s_2^x + g_3(t)\s_1^y\s_2^y].
\EEqA
In this case we use
\BEq
\a(t)=-\b(t), \quad \zeta(t)=-\xi(t),
\EEq
or
\BEqA
	U(t)= e^{-(i/2)\a(\s_1^z-\s_2^z)}
	e^{-(i/2)\left(c_1\s_1^z\s_2^z+c_2\s_1^x\s_2^x+c_3\s_1^y\s_2^y\right)}
e^{-(i/2)\zeta(\s_1^z-\s_2^z)}.
\EEqA
Eq. (\ref{eq:MAINSYSTEM2}) now becomes
\BEqA
\label{eq:MAINSYSTEM_symdetune}
\left[
\begin{array}{c}
\a'	\\
c_2' \\
c_3'\\
\zeta'\\
\end{array}
\right]=
\left[
\begin{array}{c}
\Omega_{1}
-
(g_{2}+g_{3})\cos \a \sin \a(\cos c_3 \sin c_3 - \sin c_2 \cos c_2)/
(\cos^2 c_2 - \cos^2 c_3)
\\
 g_{2} \cos^2 \a -  g_{3} \sin^2 \a \\
 g_{3} \cos^2 \a - g_{2} \sin^2 \a\\
 (g_{2}+ g_{3})\cos \a \sin \a/(\cos c_2 \sin c_3 + \sin c_2 \cos c_3)
\\
\end{array}
\right].
\EEqA

{\it Case 3.} For systems described by
\BEqA
		\label{eq:beyond tracking hamiltonian 3}
		H(t) = (1/2)[\Omega_1(t)\s_1^x+\Omega_2(t)\s_2^x+g_1(t)\s_1^x\s_2^x
		 +g_2(t)(\s_1^y\s_2^y+\s_1^z\s_2^z)]
\EEqA
use
\BEq
\label{eq:ansatz beyond 3}
c_2(t)=c_3(t), \quad \xi(t)=0,
\EEq
corresponding to
\BEqA
\label{eq:UCARTANansatz beyond 3}
U(t)= e^{-(i/2)(\a\s_1^x+\b\s_2^x)}e^{-(i/2)[c_1\s_1^x\s_2^x+
c_2(\s_1^y\s_2^y+\s_1^z\s_2^z)]} 
e^{-(i/2)\zeta \s_1^x}.
\EEqA
Notice that in this case we cannot use Eq. (\ref{eq:MAINSYSTEM2}) directly 
because matrix $M$ is not invertible, as can be seen from Eq. (\ref{eq:detM}).
Instead, the original system (\ref{eq:MAINSYSTEM1}) has to be re-written 
in accordance with 
the constraints imposed by (\ref{eq:ansatz beyond 3}). We then get
\BEqA
\label{eq:MAINSYSTEM4}
\underbrace{
\left[
\begin{array}{cccc}
1 & 0 & 0 & C_1 	\\
0 & 1 & 0 & C_2  	\\
0 & 0 & A_1 + A_2 & -(A_3-A_4)C_3 \\
0 & 0 & A_3-A_4 & (A_1+A_2)C_3 \\
\end{array}
\right]
}_{M_1}
\left[
\begin{array}{c}
\a'	\\
\b'	\\
c_2' \\
\zeta'\\
\end{array}
\right]=
\left[
\begin{array}{c}
\Omega_{1}\\
\Omega_{2}\\
g_{2}\\
0\\
\end{array}
\right],
\nonumber \\
\EEqA
with the variables defined as before, and 
\BEq
\det{M_1}=\cos c_2 \sin c_2 .
\EEq
The system can now be inverted to give
\BEqA
\label{eq: alt system}
\left[
\begin{array}{c}
\a'	\\
\b'	\\
c_2' \\
\zeta'\\
\end{array}
\right]=
\left[
\begin{array}{c}
\Omega_1 + g_2 (A_3 - A_4)\cos c_2/\sin c_2\\    
\Omega_2 + g_2 (A_3 - A_4)\sin c_2/\cos c_2\\
g_2(A_1 + A_2)\\
 -    g_2(A_3 - A_4)/\cos c_2 \sin c_2\\
\end{array}
\right],
\EEqA
which can be solved numerically.

\section{Reducing leakage to non-computational states. } 
\label{sec:REDUCING_LEAKAGE}

Here we describe two controlled-NOT gate implementations 
satisfying 
certain constraints that must be imposed on Josephson phase 
qubits \cite{MARTINIS_GELLER_private_comm} in order to make leakage 
to higher-lying (non-computational) 
states small \cite{LIDAR_PAPER1_2002, LIDAR_PAPER2_2005, 
CARLINI_PAPER2_2007}, while maintaining the high efficiency of the 
gate. The relevant conditions are:
\begin{description}
\item [1. Hamiltonian:]
\BEqA
	\label{eq:inductive coupling full}
	H(t) = (\g(t)/2)
	[\Omega_{x1}\s_1^x+\Omega_{x2}\s_2^x
	 +\Omega_{y1}\s_1^y+\Omega_{y2}\s_2^y
	 +\Omega_{z1}\s_1^z+\Omega_{z2}\s_2^z
	 +g(\s_1^x\s_2^x+\s_1^y\s_2^y+k\s_1^z\s_2^z)].
\EEqA
\item [2. Coupling constants:]
\BEqA
	\label{eq:inductive coupling constants}
	g > 0, \quad |k|<0.5.
\EEqA
\item [3. Number of $H$ applications:]
\BEq
N = 1.
\EEq
\item [4. Rabi frequencies:]
\BEqA
	\label{eq:inductive coupling Rabi}
|\Omega_{xi}|, |\Omega_{yi}|, |\Omega_{zi}| \leq g, \quad i = 1,2\,.
\EEqA
\item [5. Efficiency:]
\BEqA
	\label{eq:inductive coupling efficiency}
\eta := \frac{2\pi}{g\,t_{\rm gate}} \geq 2.5.
\EEqA
\end{description}

All these constraints can be satisfied by directly steering
toward the target belonging to the CNOT equivalence class with 
entangling part $U_{\rm ent}(t_{\rm CNOT})$ represented by the class
vector 
$\vec{c} = (\pi/2)\times (1,0,0)$. The {\it canonical} CNOT gate can 
then be made out of 
$U(t_{\rm CNOT}) = k_1U_{\rm ent}(t_{\rm CNOT}) k_2$ by
performing additional local rotations $K_1$ and $K_2$, as usual. 

The two implementations are:
\vskip5pt
\noindent 
{\bf 1. Symmetric dc detuning:} In this case the Hamiltonian is
\BEqA
	\label{eq:inductive coupling steering}
	H^{(-)}_{\rm sym.\; dc}(t) = (\g(t)/2)
	[\Omega_{1}(\s_1^x - \s_1^y) + \Omega_{2}(\s_2^x-\s_2^y)
	 +\underbrace{\Omega_{3}(\s_1^z - \s_2^z)}_{{\rm symmetric\; dc\; 
	 detuning}}
	 +g(\s_1^x\s_2^x+\s_1^y\s_2^y+k\s_1^z\s_2^z)],
\EEqA
or, alternatively,
\BEqA
	\label{eq:inductive coupling steering alt}
	H^{(+)}_{\rm sym.\; dc}(t) = (\g(t)/2)
	[\Omega_{1}(\s_1^x + \s_1^y) + \Omega_{2}(\s_2^x + \s_2^y)
	  +\underbrace{\Omega_{3}(\s_1^z - \s_2^z)}_{{\rm symmetric\; dc\; 
	  detuning}}
	 +g(\s_1^x\s_2^x+\s_1^y\s_2^y+k\s_1^z\s_2^z)],
\EEqA
with
\BEq
 \Omega_{1} \equiv g,
\EEq
where $\g(t)$ represents experimentally available tracking control, and the 
superscript $^{(\pm)}$ refers to the corresponding choice of the $x$ and $y$ 
Rabi parts.
The relevant control parameters have been found numerically and are listed 
in Table \ref{tab:table2}.

\begin{table}
\caption{
\label{tab:table2} Generation of controlled-NOT logic 
with $\vec{c} = (\pi/2)\times (1,0,0)$ using inductively coupled flux qubits 
driven by weak local perturbations and subject to symmetric dc detuning 
$(\Omega_{3}/2)(\s_1^z - \s_2^z)$. The Hamiltonian is given in  
Eq. (\ref{eq:inductive coupling steering}) or 
(\ref{eq:inductive coupling steering alt}), where $\Omega_1/g \equiv 1$. Here, 
$\g(t)=1$.
}
\begin{ruledtabular}
\begin{tabular}{ccccc}
$k$ & $t_{\rm CNOT}$, units of $\pi/2g$ & $\Omega_2^{\rm CNOT}/g$ & 
$\Omega_3^{\rm CNOT}/g$  & $\eta$ \\ \hline
 0.000 	& 1.595776  & 0.000000 &  0.755502	&	2.5066\\ 
 0.001 	& 1.595775  & 0.000264 &  0.755503	& 2.5066\\ 
 0.002 	& 1.595774  & 0.000529 &  0.755505 	& 2.5066\\
 0.003 	& 1.595772  & 0.000793 &  0.755509	& 2.5066\\ 
 0.004 	& 1.595769  & 0.001057 &  0.755515	& 2.5066\\
 0.005	& 1.595765  & 0.001322 &  0.755522	& 2.5066\\ \hline
 
 0.010	& 1.595731  & 0.002644 &  0.755582  & 2.5067\\ 
 0.025	& 1.595496  & 0.006614 &  0.756001	& 2.5071\\
 0.050	& 1.594657  & 0.013257 &  0.757500	& 2.5084\\ 
 0.075	& 1.593263  & 0.019961 &  0.760001	& 2.5106\\ 
 0.100	& 1.591321  & 0.026758 &  0.763506	& 2.5136\\ \hline
 
 0.150  & 1.585843  & 0.040779 &  0.773549  & 2.5223\\
 0.250	& 1.569080  & 0.071908 &  0.806036	& 2.5493\\
 0.350	& 1.547002  & 0.111865 &  0.856120	& 2.5856\\ 
 0.450	& 1.530753  & 0.178169 &  0.927506	& 2.6131\\ \hline
 0.490	& 1.550430  & 0.240369 &  0.966790	& 2.5799\\ 
 0.493  & 1.561200  & 0.254105 &  0.971189  & 2.5621 \\
  \end{tabular}
\end{ruledtabular}
\end{table}

\begin{figure}
\includegraphics{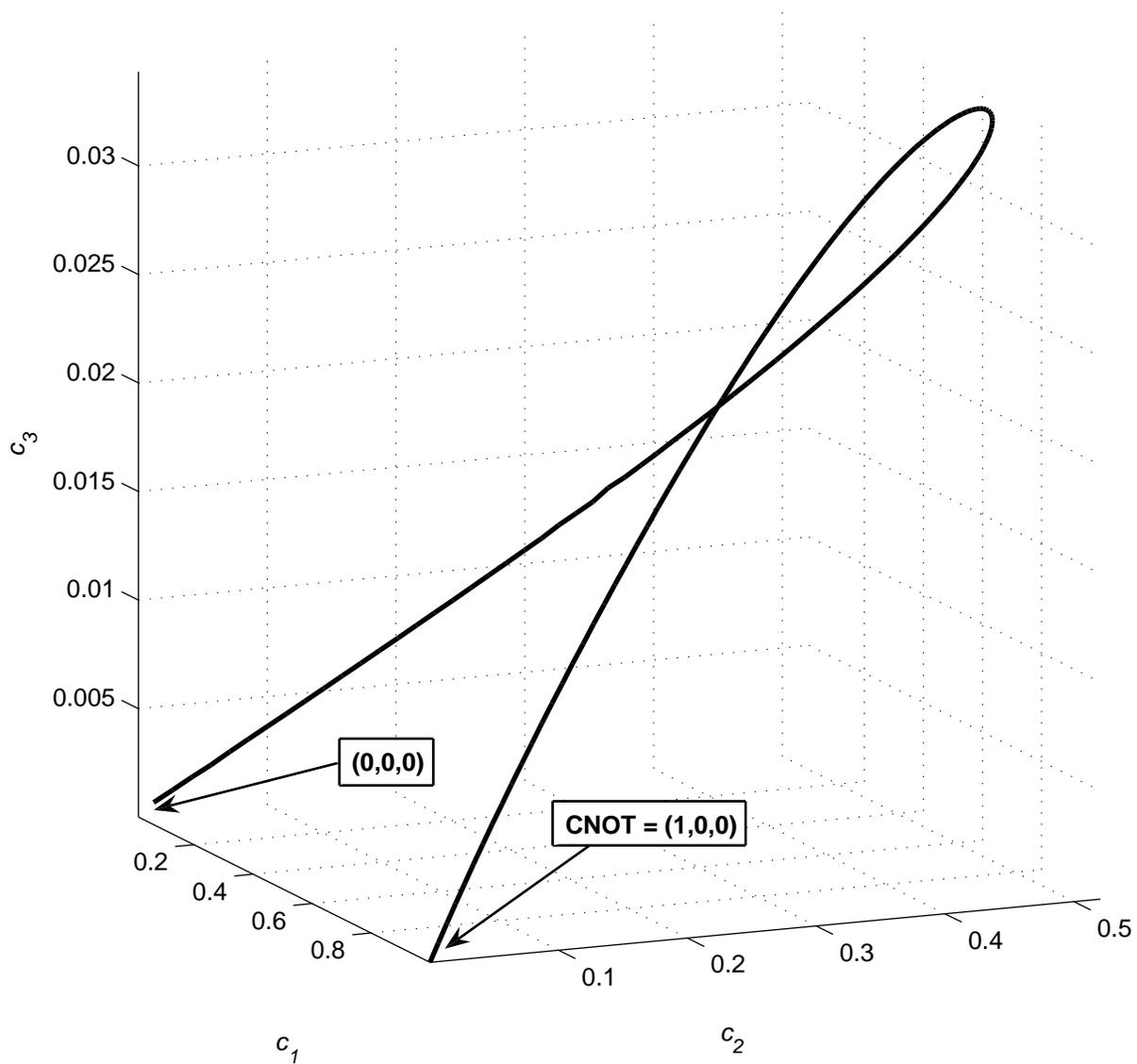}
\caption{
\label{fig:6} 
Weyl chamber steering trajectory generating CNOT 
class in the case of inductively coupled
flux qubits driven by weak local perturbations and subject to symmetric 
dc detuning, Eq. (\ref{eq:inductive coupling steering}). Here, $g=1.00$, 
$k=0.050$, $\Omega_2 = 0.0133$,   $\Omega_3 = 0.7575$, $\g(t)=1$. 
The steering parameters are given in units of $\pi/2$.
}
\end{figure}

\begin{figure}
\includegraphics{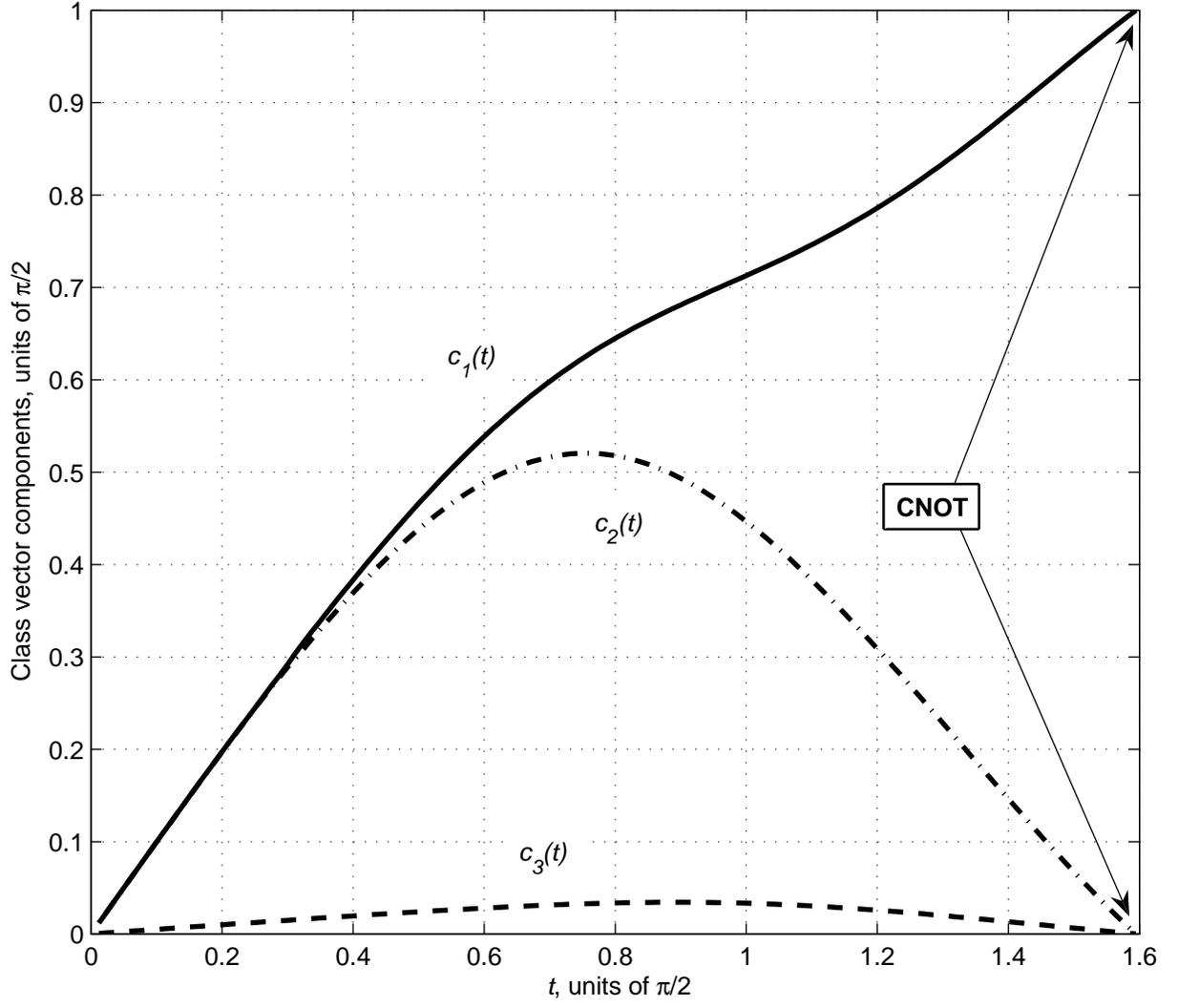}
\caption{
\label{fig:7} 
Time dependence of Weyl chamber steering 
parameters shown in Fig. \ref{fig:6}.
}
\end{figure}

Figures \ref{fig:6} and \ref{fig:7} show the steering trajectory for 
$g=1.00$, $k=0.050$, with Rabi frequencies $\Omega_2 = 0.0133$, 
$\Omega_3 = 0.7575$.

\vskip5pt
\noindent 
{\bf 2. Asymmetric dc detuning:} In this case the Hamiltonian is
\BEqA
	\label{eq:inductive coupling steering asym}
	H^{(-)}_{\rm asym.\; dc}(t) = (\g(t)/2)
	[\Omega_{1}(\s_1^x - \s_1^y) + \Omega_{2}(\s_2^x-\s_2^y)
	  +\underbrace{(\Omega_{3}\s_1^z - \Omega_{4}\s_2^z)}_{{\rm asymmetric\; 
	 dc\; detuning}}
	   +g(\s_1^x\s_2^x+\s_1^y\s_2^y+k\s_1^z\s_2^z)],
\EEqA
or, alternatively,
\BEqA
	\label{eq:inductive coupling steering asym alt}
	H^{(+)}_{\rm asym.\; dc}(t) = (\g(t)/2)
	[\Omega_{1}(\s_1^x + \s_1^y) + \Omega_{2}(\s_2^x + \s_2^y)
	  +\underbrace{(\Omega_{3}\s_1^z - \Omega_{4}\s_2^z)}_{{\rm asymmetric\; 
	 dc\; detuning}}
	  +g(\s_1^x\s_2^x+\s_1^y\s_2^y+k\s_1^z\s_2^z)],
\EEqA
with
\BEq
 \Omega_{1} = \Omega_{3} \equiv g.
\EEq
The corresponding steering controls are listed in Table \ref{tab:table3}.

\begin{table}
\caption{
\label{tab:table3} 
Generation of controlled-NOT 
logic with $\vec{c} = (\pi/2)\times (1,0,0)$ using inductively coupled flux 
qubits driven by weak local perturbations and subject to asymmetric dc 
detuning
$(\Omega_{3}/2)\s_1^z - (\Omega_{4}/2)\s_2^z$. The Hamiltonian is given 
in Eq. (\ref{eq:inductive coupling steering asym}) or 
(\ref{eq:inductive coupling steering asym alt}), where 
$\Omega_1/g = \Omega_3/g \equiv 1$. Here, $\g(t)=1$.
}
\begin{ruledtabular}
\begin{tabular}{ccccc}
$k$ & $t_{\rm CNOT}$, units of $\pi/2g$ & $\Omega_2^{\rm CNOT}/g$ & 
$\Omega_4^{\rm CNOT}/g$  & $\eta$ \\ \hline
 0.000 	& 1.553771 &  0.000000 &  0.402539	& 2.5744	\\ 
 0.001 	& 1.553770 &  0.000179 &  0.402541	& 2.5744\\ 
 0.002 	& 1.553768 &  0.000358 &  0.402548	& 2.5744\\
 0.003 	& 1.553766 &  0.000537 &  0.402558  & 2.5744\\ 
 0.004 	& 1.553762 &  0.000715 &  0.402574	& 2.5744\\
 0.005	& 1.553757 &  0.000894 &  0.402593	& 2.5744\\ \hline
 
 0.010	& 1.553716 &  0.001789 &  0.402757  & 2.5745\\ 
 0.025	& 1.553430 &  0.004475 &  0.403902	& 2.5749\\
 0.050	& 1.552414 &  0.008974 &  0.407988	& 2.5766\\ 
 0.075	& 1.550736 &  0.013523 &  0.414781	& 2.5794\\ 
 0.100	& 1.548418 &  0.018150 &  0.424259	& 2.5833\\ \hline
 
 0.150  & 1.541995 &  0.027780 &  0.451143  & 2.5940\\
 0.250	& 1.523410 &  0.050016 &  0.535559	& 2.6256\\
 0.350	& 1.501442 &  0.081649 &  0.659439	& 2.6641\\ 
 0.450	& 1.488962 &  0.141937 &  0.826279	& 2.6864\\ \hline 
 0.500  & 1.515587 &  0.220268 &  0.938373  & 2.6392\\
 0.506  & 1.539498 &  0.251771 &  0.959755  & 2.5982\\
 \end{tabular}
\end{ruledtabular}
\end{table}

\section{Conclusion}

In summary, we have proposed a self-contained approach to steering on 
the Weyl chamber and applied it to the case of anisotropic 
exchange with tracking controls, which was solved analytically. It was 
shown that if architecture allows for local manipulation of individual 
qubits, {\it any exchange interaction can generate CNOT quantum logic}. 
The 
results were then
used to identify several CNOT gate implementations for
superconducting Josephson qubits, including the ones that are capable 
of suppressing leakage to noncomputational states without significant 
reduction in the gate's efficiency.

\begin{acknowledgments}

This work was supported by the Disruptive Technology Office under
Grant No. W911NF-04-1-0204 and by the National Science Foundation under
Grant No. CMS-0404031. The author thanks Michael Geller, John Martinis, 
Emily Pritchett, 
and Andrew Sornborger for useful discussions.

\end{acknowledgments}

\newpage

\end{document}